\documentclass[aps,prb,showpacs,floatfix,twocolumn,superscriptaddress]{revtex4-2}
\usepackage{graphicx}
\usepackage{dcolumn}
\usepackage{cancel}
\usepackage{bm}
\usepackage{tikz}
\usepackage{siunitx}
\usepackage{tablefootnote}
\usepackage{physics}
\usepackage[section]{placeins}
\definecolor{darkblue}{rgb}{0,0,.65}
\definecolor{darkgreen}{rgb}{1,0,0}
\usepackage{nicefrac}
\usepackage{booktabs}
\usepackage[%
  pdfauthor={Benedikt Placke},%
  pdfstartview=FitH,%
  breaklinks=true,%
  bookmarks=true,%
  colorlinks=true,%
  anchorcolor=black,%
  citecolor=red,
  filecolor=black,%
  menucolor=black,%
  urlcolor=blue,%
  linkcolor=red,%
 ]{hyperref}

\newcommand{\s}{\mathbf{s}}
\usepackage[english]{babel}
\usepackage{bm}
\usepackage{amssymb}
\usepackage{color, soul}
\usepackage{amsmath}
\usepackage{amstext}
\usepackage{latexsym}
\usepackage{graphicx}
\usepackage{mathtools}
\usepackage[ruled]{algorithm2e}

\begin{document}
\title{Tensor network Monte Carlo simulations for the two-dimensional random bond Ising model}

\author{Tao Chen}%
\affiliation{
Hefei National Laboratory for Physical Sciences at the Microscale and Department of Modern Physics, University of Science and Technology of China, Hefei 230026, China}
\affiliation{Hefei National Laboratory, University of Science and Technology of China, Hefei 230088, China}

\author{Erdong Guo}
\affiliation{
CAS Key Laboratory for Theoretical Physics, Institute of Theoretical Physics,
 Chinese Academy of Sciences, Beijing 100190, China}

\author{Wanzhou Zhang}
\thanks{zhangwanzhou@tyut.edu.cn}
\affiliation{College of Physics, Taiyuan University of Technology, Shanxi 030024, China}


\author{Pan Zhang}%
\thanks{panzhang@itp.ac.cn}
\affiliation{Hefei National Laboratory, University of Science and Technology of China, Hefei 230088, China}
\affiliation{
CAS Key Laboratory for Theoretical Physics, Institute of Theoretical Physics,
 Chinese Academy of Sciences, Beijing 100190, China}
\affiliation{
School of Fundamental Physics and Mathematical Sciences, Hangzhou Institute for Advanced Study, UCAS, Hangzhou 310024, China
}

\author{Youjin Deng}%
\thanks{yjdeng@ustc.edu.cn}
\affiliation{
Hefei National Laboratory for Physical Sciences at the Microscale and Department of Modern Physics, University of Science and Technology of China, Hefei 230026, China}
\affiliation{Hefei National Laboratory, University of Science and Technology of China, Hefei 230088, China}

\date{\today}

\begin{abstract}

Disordered lattice spin systems are crucial in both theoretical and applied physics. However, understanding their properties poses significant challenges 
{to} Monte Carlo simulations. In this work, we investigate the two-dimensional {random bond} Ising model using the recently proposed Tensor Network Monte Carlo (TNMC) method. This method generates biased samples from conditional probabilities computed via tensor network contractions and corrects the bias using the Metropolis scheme. Consequently, the proposals provided by tensor networks function as block updates for Monte Carlo simulations.
Through extensive numerical experiments, we demonstrate that TNMC simulations can be performed on lattices as large as \(1024 \times 1024\) spins with moderate computational resources, a substantial increase from the previous maximum size of \(64 \times 64\) in MCMC. Notably, we observe an almost complete absence of critical slowing down, enabling the efficient collection of unbiased samples and averaging over a large number of random realizations of bond disorders.
We successfully pinpoint the multi-critical point along the Nishimori line with significant precision and accurately determined the bulk and surface critical exponents. Our findings suggest that TNMC is a highly efficient algorithm for exploring disordered and frustrated systems in two dimensions.
\end{abstract}

\maketitle
\section{Introduction}
Studying lattice spin systems is of great importance in various fields of physics and materials science due to their rich theoretical and practical implications. However, the intrinsic high dimensionality makes the study challenging, especially in systems where interactions compete and exhibit disorders, such as in spin glasses or geometrically frustrated lattices.
Developing efficient, accurate, and scalable algorithms is still an urgent and demanding task. 
The Markov Chain Monte Carlo (MCMC) method~\cite{Metropolis1953EquationOS,landau_binder_2014} is one of the most widely used and efficient methods for studying lattice spin systems, especially classical ones. However, they {meet} several challenges, 
including the dramatic growth of the autocorrelation time 
near critical points and at low temperatures, termed the 
``critical slowing down''~\cite{csd}, 
especially in systems with rugged energy landscapes~\cite{Edwards1975, binder1986}. 
Such systems (e.g. spin glasses) have many metastable states, and local update MCMC algorithms (e.g., Metropolis-Hastings) 
can easily get trapped in {the} local minima. This makes it difficult to sample the configuration space effectively. 
As a concrete example, for the $L\times L$ two-dimensional (2D) {random bond Ising model}, 
even enhanced with the parallel tempering technique, 
extensive MCMC simulations with {the} {Metropolis-Hasting} algorithm were reported 
only for systems with maximum size $L=64$~\cite{MC2008,MC2009}.
In 2D and at zero temperature, thanks to insights from graph theory, 
efficient algorithms have been developed to sample ground states for the random field 
and random bond Ising models~\cite{acm,PhysRevLett.116.227201,PhysRevE.93.063308,Amoruso-domain-wall}.

A groundbreaking progress in suppressing critical slowing down  was 
the Swendsen-Wang~\cite{PhysRevLett.58.86} and the Wolff~\cite{WOLFF199093} cluster algorithms. 
The Ising and the Potts models take the Fortuin-Kasteleyn representation and propose to flip an entire cluster of spins. The updating of clusters of spins leads to a significant change of configuration compared with local updates and effectively explores the configuration space.
However, the cluster updates in the Swendersen-Wang and the Wolff algorithms are highly specific 
to certain problems and not generally applicable. 
In the disordered systems, the clusters identified by the cluster algorithms
are typically either too large or too small, which highly reduces its efficiency, making the cluster-update {proposal perform even worse} than the {Metropolis} single-spin update.

Tensor network methods~\cite{nishino,dmrg,dmrg_age,nishino3d,trg_1,qiaoni_chen,strg,CTMRG,tropical,orus,sjran,rmp,Xiang_2023} 
have been widely used in lattice spin systems, offering an accurate way of estimating properties 
such as the partition function, magnetizations and correlations. 
One can also generate samples efficiently. 
However, the truncation errors resulting from the singular value decompositions (SVD) 
make the estimates and the samples intrinsically biased, 
as opposed to MCMC which is numerically exact and gives unbiased estimates. 
Recently, efforts have been devoted to combining the Monte Carlo and the tensor network techniques.
In Refs.~\cite{random_svd,ferris2015unbiased}, Monte Carlo sampling was introduced 
in the contraction of tensor networks to obtain numerically exact partition function.
In Ref.~\cite{tnmc}, approximation in the contration process was kept unchanged,
and the MCMC method was then introduced to globally update spin configurations 
based on the approximate partition function.
Refs.~\cite{random_svd,ferris2015unbiased} aimed at using sampling in order to estimate a tensor 
network contraction, whereas Ref.~\cite{tnmc} was  to explore the `dual' 
idea: using tensor network contraction for sampling.
More precisely speaking, instead of constructing random clusters, 
the tensor network Monte Carlo (TNMC) method utilizes tensor network contractions 
to approximately compute the joint distribution and conditional distributions of 
a chosen block of spins, which can correspond to 
the total or part of the physical system, and generate samples according to the conditional probabilities. 
The samples can be treated as proposals and further accepted or rejected according to 
the Metropolis acceptance-rejection scheme to generate unbiased samples for the physical system. 
TNMC corrects the bias of the tensor network using the Metropolis scheme 
and introduces the block updates to MCMC using the tensor network computation, 
thereby addressing issues inherent in both tensor networks and Monte Carlo (MC) methods.
The rapid growth of the autocorrelation time in conventional MC algorithms 
can be dramatically suppressed, 
since the proposed configuration from the tensor network 
is global and is normally accepted with significant probability. 
{Conversely}, since the bias is corrected by the Metropolis scheme, 
the tensor network computation can adopt a small bond dimension and consume moderate computation resources.
Despite its potential, recent studies on TNMC are still primarily demonstrative and proof-of-concept.

In this work, we extend the TNMC method in 
Ref.~\cite{tnmc} to simulate the 2D {random bond} Ising model along the Nishimori line,
which is challenging for MCMC. 
We show that with moderate computational resources, 
one can achieve simulation on lattices with $1024\times 1024$ spins, 
which is significantly greater than the previous MCMC study with $64\times 64$ spins~\cite{MC2008,MC2009}. 
Even on such large systems, we observe almost no critical slowing down, this allows the method 
to generate efficiently the {uncorrelated} samples. 
This suggests that the TNMC method can serve as a powerful tool for 
studying disordered systems at least in two dimensions, 
without the need of the broadly used parallel tempering.
We then locate the multi-critical point along the Nishimori line with high-precision 
and accurately determine the bulk and surface critical exponents.
Taking into account that the TNMC method is a hybrid algorithm combining MCMC and 
tensor network techniques and much details are not discussed in Ref.~\cite{tnmc}, 
we illustrate the basic ideas in Appendix in the example 
of the one-dimensional Ising model, and, further, have made our codes {publicly} 
available on {Github}~\cite{github} for the convenience of readers.

The rest of the paper is organized as follows.
In Sec.~\ref{sec:methods}, 
we introduce the TNMC method and the block update. In Sec.~\ref{sec:results}, we simulate the {random bond} $\pm J$ Ising model {and} present the simulation results.
Conclusions are made in Sec.~\ref{sec:conclusion}.

\section{Tensor Network Monte Carlo}
\label{sec:methods}
Consider the Boltzmann distribution of the Ising model
\begin{equation}\label{eq:P}
P(\mathbf s)=\frac{1}{Z}e^{-\beta E(\mathbf s)},
\end{equation}
where $\mathbf s\in\{+1,-1\}^N$ is a configuration of $N$ spins, $\beta$ is the inverse temperature, $E(\mathbf s)$ is the energy function and 
\begin{equation} \label{eq:Z}
Z=\sum_{\mathbf s}e^{-\beta E(\mathbf s)}
\end{equation} 
is the partition function. 
In this work, we investigate the {random bond} Ising model on the square lattice,
and free boundary conditions are considered 
along both the $x$ and $y$ directions. 
The MCMC method constructs a Markov chain with the target Boltzmann distribution as its equilibrium distribution. 
The properties of the Boltzmann distribution can be estimated using samples drawn from the distribution 
once the Markov chain has sufficiently mixed. 
Giving a current configuration $\s_a$, a candidate configuration $\s_b$ is generated from a proposal distribution $g(\s_b|\s_a)$, and is accepted with probability 
\begin{eqnarray}    p_a(\s_b|\s_a)=\min\left\{1,\frac{g(\s_a|\s_b)}{g(\s_b|\s_a)}\times\frac{P(\s_b)}{P(\s_a)}\right\},
\label{eq:accptanceratio}
\end{eqnarray}
which is called the Metropolis filter or the Metropolis acceptance-rejection scheme. 
In lattice spin systems, {to ensure that} the acceptance probability {is} not much smaller than $1$, new configurations that are close to the current configurations with similar energy are often chosen, known as local moves.
The Metropolis-Hasting method is one of the most widely used MCMC algorithms, which chooses a candidate with only one spin flipped from the current configuration. This is because the acceptance rate of a configuration is exponential in its energy difference $$\frac{P(\s_b)}{P(\s_a)}=e^{\beta E(\s_a)-\beta E(\s_b)}.$$ 
{Therefore}, candidates which are far from the current configuration have a high probability of being rejected.
As a consequence, the subsequent configurations in the Metropolis-Hasting algorithm typically have a strong correlation, which leads to severe critical slowing down. 
In particular, the local moves can get trapped in the local minima, 
especially in disordered systems, because due to the nature of the rugged energy landscape, 
the probability of moving to a higher energy state is low. 
This makes the Markov chain hard to mix and makes MCMC difficult to accurately sample the equilibrium distribution. 

\subsection{Tensor network {proposals}}
To overcome the issue of local updates, in~\cite{random_svd} the tensor networks are utilized for generating global moves, i.e. proposal configurations that are much different from the current configuration.
This is achieved by sampling from the Boltzmann distribution using conditional probabilities as follows:
\begin{equation}\label{eq:Bayes}
P(\s) = \prod_{i=1}^NP(s_i|\mathbf {s}_{<i}).
\end{equation}
In the above equation, $P(s_i|\mathbf {s_{<i}})$ is the conditional probability of 
spin $i$ given the configuration of spins in front of $i$.
An {illustrative} example using the one-dimensional (1D) Ising model 
is given in detail in Appendix~\ref{sec:1d}. 
For a 2D lattice, one can choose a natural spin order such as a Zig-Zag order, as illustrated in Fig.~\ref{fig:2d_tn}. 
The conditional probabilities can be computed as:
\begin{equation}\label{eq:cond}
P(s_i|\mathbf {s_{<i}}) = \frac{\sum_{\s_{>i}}e^{-\beta E(s_i,\s_{<i})}}{\sum_{s_i,\s_{>i}}e^{-\beta E(s_i,\s_{<i})}}=\frac{Z(s_i,\s_{<i})}{\sum_{s_i}Z(s_i,\s_{<i})}.
\end{equation}
Here a configuration $\mathbf s = \{\mathbf s_{<i},s_i,\mathbf s_{>i}\}$ is separated into three parts. 
The configuration of spins behind $i$, $\mathbf s_{>i}$, is summed over, 
and the configuration of spins in front of $i$, $\mathbf s_{<i}$, is given;
{the} decomposition of the Boltzmann distribution in Eq.~\eqref{eq:Bayes} is 
nothing more than utilizing the Bayes rule and hence is exact. 
If one can compute the conditional probabilities exactly, this results in an unbiased sampling of the Boltzmann distribution, with an acceptance probability of 1. Consequently, the observables estimated from the samples are unbiased.
However, the computation of the conditional probabilities requires evaluating 
$Z(s_i,\s_{<i})$, the conditional partition function where the spins $\s_{<i}$ are explicitly given. This calculation belongs to the computational class of \#P-Hard problems 
and no polynomial algorithm can achieve the exact computation for general problems. 

\subsection{Computing the partition function using tensor networks} 
Tensor renormalization group methods have been widely used for computing partition functions of lattice spin models~\cite{nishino,dmrg,dmrg_age,nishino3d,trg_1,qiaoni_chen,strg,CTMRG,tropical,orus,sjran,rmp,Xiang_2023}, 
by converting the summing over an exponential number of configurations using tensor network contractions. 
In TNMC~\cite{random_svd}, the tensor network is utilized to approximately compute the conditional partition functions $Z(s_i,\s_{<i})$ 
and the conditional probabilities $P(s_i|\mathbf {s_{<i}})$.
For 1D lattices, the tensor contraction for computing the conditional probabilities can be made exact, 
we give an illustrated example in the Appendices. 
For 2D systems, tensor network contractions must evolve compression of tensors 
(i.e. small bond dimensions in the language of the tensor networks) and 
give approximate conditional probabilities $q(s_i|\mathbf s_{<i})$. This results in an approximation to the Boltzmann distribution 
\begin{equation}\label{approximate_bayes}
q(\mathbf s) = \prod_{i=1}^Nq(s_i|\mathbf {s_{<i}}).
\end{equation}

\begin{figure}[h]
\includegraphics[width = 0.9\linewidth]{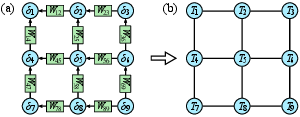}
\caption{
(a) The tensor network corresponding to the partition function computation of the Ising model on the $3\times 3$ lattice. $\delta$ denotes the copy tensor associated with each spin and $W$ denotes the {Boltzmann} matrices associated with each coupling (see text). 
(b) The $T$ tensor is obtained by contracting the copy tensor with two $W$ matrices on the bottom and on the right.}
\label{fig:2d_tn}
\end{figure}
For the convenience of readers, as an illustrative example, we use a $3 \times 3$ square lattice 
with free boundary conditions
to illustrate the construction and the contraction of the tensor network
and the sampling of the Boltzmann distribution, taking the Ising model as an example.
As shown in Fig.~\ref{fig:2d_tn}, the spins are ordered 
from the top left to the bottom right of the lattice, 
with labels $i=1,2,\cdots, 9$ respectively.
In the tensor network, {each} spin on the lattice site corresponds to a $m$-leg \textit{copy tensor} $\delta$,
where the number of legs can be $m=2,3$, or 4, 
depending on whether the site is on the corners, on the edges, or in the center. 
The elements of $\delta$ are equal to 1 if all of the tensor indices are identical; otherwise, they are 0. 
For instance, a 4-leg \textit{copy tensor} reads as:
\begin{equation}
    \delta_{iabl}=\left\{
    \begin{aligned}
        1&\quad i=a=b=l \\
        0&\quad {\rm else} \quad (i,a,b,l=1,2)
    \end{aligned}
    \right. \; .
\end{equation}
The elements, $\delta_{1111}$ and $\delta_{2222}$, {mean} 
the Ising spin {take the values} $+1$ and $-1$, respectively. 
Further, to represent a fixed Ising spin $s$, 
a reduced rank-one tensor $\delta'$ can be introduced such that 
only a single element is nonzero: $\delta_{1111}=1$ if $s=1$ 
or $\delta_{2222}=1$ if $s=-1$.
On each interacting edge between two neighboring sites $i$ and $j$, 
there is a Boltzmann matrix in the tensor network representation
\begin{equation}\label{BoltzmannMatrix}
W_{ij} =
    \begin{pmatrix}
    e^{\beta J_{ij}} & e^{-\beta J_{ij}} \\
    e^{-\beta J_{ij}} & e^{\beta J_{ij}}
    \end{pmatrix} \; ,
\end{equation}
where $W_{11}=W_{22}=\exp(\beta J_{ij})$ corresponds to a pair of parallel Ising spins and $W_{12}=W_{21}=\exp(-\beta J_{ij})$
corresponds to a pair of anti-parallel spins, with subscripts ``1'' and ``2'' 
for spins $s=+1$ and $s=-1$, respectively. 

\begin{figure}[h]
\includegraphics[width = 0.85\linewidth]{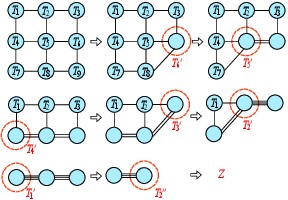}
\caption{Illustration of the tensor network contraction process. 
The blue circles labeled with $T_i$ represent the original tensors, 
and the unlabeled blue circles indicate {higher order} tensors 
obtained after the contraction of neighboring tensors. 
The contraction proceeds sequentially from $T_9$ to $T_1$
{(see text)},
and, by contracting all the tensors, the total partition function $Z$ is obtained. 
During the contraction process, the tensors enclosed by the red dashed lines 
are the process tensors involved at each step, and they are stored in computer memory
for later use in the sampling process. }
\label{fig:contraction}
\end{figure}
With the tensors $W$ and $\delta$, the partition function $Z$ in Eq.~(\ref{eq:Z}) 
can then be expressed in a network of tensors, as shown in Fig.~\ref{fig:2d_tn}(a). 
{In} the first step to compute $Z$, one can contract the copy tensors $\delta$ with the $W$ matrices 
on the right and the bottom, giving a network of $T$ tensors as shown in Fig.~\ref{fig:2d_tn}(b)
\begin{equation}
T_{ijlk}=\sum_{a=1}^2\sum_{b=1}^2\delta_{iabl}W_{aj}W_{bk} \;.
\end{equation}

Next, we sequentially contract the $T$ tensors starting from the last one $T_9$ 
and repeats the process till the first one $T_1$, as illustrated in Fig.~\ref{fig:contraction}.
During the contraction process illustrated in Fig.~\ref{fig:contraction}, 
a series of process tensors are obtained, represented by the tensor enclosed within the red dashed lines. 
These process tensors are stored to calculate the conditional partition functions in 
the sampling process. 
The total partition function $Z$ is obtained after contracting the final two process tensors $T_2''$ and $T_1'$. 
Throughout the contraction process, it is crucial to cache both the original tensors and the process tensors, in order to sample spin configurations using the conditional probabilities. Moreover, identical process tensors can be cached once to optimize storage efficiency.

\begin{figure}[h]
\includegraphics[width = 0.85\linewidth]{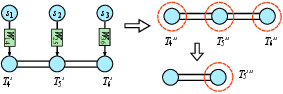}
\caption{Illustration of the preparation for sampling the spins in the second row. 
After the first three spins are fixed as $\s_{<4}=\{s_1,s_2,s_3\}$, 
the corresponding copy tensors $\delta$ on these lattice sites 
are replaced by the reduced copy tensors $\delta'$ representing the fixed spin value. 
The interaction matrices connected to the next row are 
then contracted as external fields into the process tensors, 
resulting in new process tensors $T_4'',T_5''$, and $T_6''$. 
The contraction process then proceeds from right to left, using cached process tensors enclosed within the red dashed circles.}
\label{fig:Z4}
\end{figure}

\begin{figure}[h]
\includegraphics[width = 0.85\linewidth]{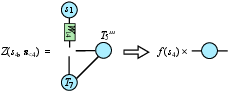}
\caption{Illustration of computation of the conditional partition function $Z(s_4,\s_{<4})$.
The final result is a $2\times 2$ matrix combined with the function $f(s_4)=e^{\beta J_{14}s_1s_4}$, in which the (1,1)$\times f$(1) element 
is the conditional partition function for $s_4=+1$ and 
the (2,2)$\times f$(-1) element is the conditional partition function for $s_4=-1$. }
\label{fig:ZS4}
\end{figure}
During the contraction of tensors, the bond dimension $D$ of the intermediate tensors increases rapidly. 
In the actual computation, the bond dimension is truncated to a fixed maximum value 
by using the {singular value decomposition} (SVD), which makes an approximation to the computation.
We refer to ~\cite{ferris2015unbiased,tnmc} for details of the obtained conditional partition functions and the corresponding conditional probabilities.

The sampling process follows the reverse order of the tensor network contraction. 
For example, to sample spin $s_i$ ($i=1,\cdots,9$), we first evaluate 
the conditional partition functions $Z(s_i,\s_{<i})$ in Eq.~(\ref{eq:cond}). 
This is computed by contracting the process tensor $\{T_i'\}$ with the reduced tensor with sampled spin configuration $\s_{<i}$.
Here we take the sampling process for $s_4$ as an illustrative example. The spin configuration for $s_1,s_2$, and $s_3$ {is} determined (i.e. sampled already), the corresponding copy tensors $\delta$ (in Fig.~\ref{fig:2d_tn}) reduce to $\delta'$ according to their configuration $s_i$ (in Fig.~\ref{fig:Z4}). 
Since there is only one single nonzero element in $\delta'$,
the combined effect of the reduced copy tensors $\delta'$ and the 
Boltzmann matrix $W_{14},W_{25},W_{36}$ can be regarded as 
the external fields applied to the process tensors, 
resulting in the tensors $T_4'',T_5''$ and $T_6''$. 
We then contract the process tensors from the right to the left 
and obtain the process tensor $T_5'''$. 
The values of the new process tensors depend on the previously sampled 
spins $\{s_1,s_2,s_3\}$, and they can be cached for future use.
Finally, we contract the process tensor $T_5'''$ with the original tensor $T_7$ to obtain the condition partition function $Z(s_4,\s_{<4})$ as shown in Fig.~\ref{fig:ZS4}. This incorporates the function $f(s_4)=e^{\beta J_{14}s_1s_4}$ which represents the interaction between $s_1$ and $s_4$. 
Then the conditional probability 
$P(s_4|\s_{<4})$ is evaluated using Eq.~(\ref{eq:cond}). 
The other spins are sampled similarly. Notice that the sampling process allows a large number of configurations to be sampled {in parallel} and the configurations generated in this way are independent. 
If the sampled configurations can be accepted with high probabilities, the 
{autocorrelation} issue of MCMC can be solved. 
It has been demonstrated that the samples are unbiased~\cite{random_svd,ferris2015unbiased}, and gives an accurate estimate of the partition function for 2D ferromagnetic Ising model~\cite{random_svd}. 
Moreover, the acceptance ratios are high even with relatively small bond dimensions for 2D 
ferromagnetic systems with $256\times256$ spins and frustrated systems with $128\times 128$~\cite{tnmc}.
Finally, it is noted that our contraction and sampling procedures 
differ slightly from those in Ref.~\cite{tnmc}, 
to improve the clarity of physical meanings as well as the computation efficiency.

\section{Results}
\label{sec:results}

In this section, we apply the TNMC method to investigate the $\pm J$ 
{random bond} Ising model 
on the square lattice with free boundary conditions. 

\subsection{Random-bond Ising model}
The energy of the $\pm J$ {random bond} model is given by
\begin{equation}\label{eq:E}
E(\mathbf s)=-\sum_{(ij)}J_{ij}s_is_j,
\end{equation}
where the summation is over the nearest neighbors 
and $J_{ij}$ is taken from $\{+1,-1\}$ randomly, with 
\begin{equation}
\mathcal{P}\left(J_{ij}\right)=(1-p)\delta\left(J_{ij}-1\right)+p\delta\left(J_{i j}+1\right),
\end{equation}
where $p$ is the probability of anti-ferromagnetic bonds.

\begin{figure}[t]
\centering
\includegraphics[width =0.80\linewidth]{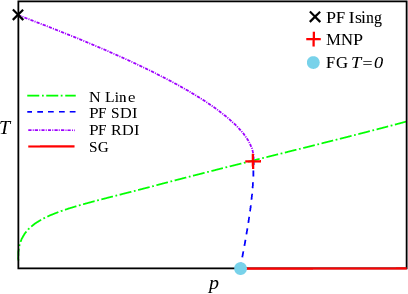}
\caption{The schematic phase diagram of the square-lattice $\pm J$ 
{random bond}  Ising model. 
The phase diagram contains ferromagnetic (F), paramagnetic (P), and spin glass (SG) phases.
The meanings of symbols are: ``PF Ising'' for pure Ising, ``MNP'' for multi-critical Nishimori point, 
``FG $T=0$'' for zero-temperature ferromagnet-glass transition, 
``N line'' for Nishimori line, ``PF SDI''for paramagnet-ferromagnet strong disordered Ising, 
and ``PF-RDI'' for paramagnet-ferromagnet randomly diluted Ising. }
\label{fig:phase2}
\end{figure}

\begin{table*}[t]
\centering
\caption{Brief summary of the estimates of 
the multi-critical Nishimori point and the critical exponents 
for the $\pm J$ random bond Ising model on the square lattice. 
Here, $L_{\rm max}$ represents the largest system size in each work, 
$p_c$ is the critical probability for antiferromagnetic interaction, 
$y_1$ is the inverse correlation length exponent, 
and $\eta$ and $\eta_{\text{edge}}$ are for the bulk and the 
surface (edge) magnetic anomalous dimensions, respectively.
}\label{tablelist}
\vskip 0.2cm
\begin{ruledtabular}
\begin{tabular}{lclllllc}
Methods                       & Date       & $L_{\rm max}$ & $p_c<1/2$          & $y_1$         & $\eta$          & $\eta_{\text{edge}}$                \\
\hline      
             Transfer Matrix (TM)~\cite{tm1987}           & 1987       & 14        & 0.111(2)           & -       &  -                 \\
             Monte Carlo Renormalization Group (MCRG)~\cite{mcrg}           & 1987       & 64        & 0.11(1)           & -       &  -                    \\
             TM~\cite{Ueno1991}           & 1991       & -         & -           & -       & -                  \\
            Expansion~\cite{pc5}           & 1996       & -         & 0.114(3)           & 0.75(7)       & 0.20(1)                       \\
          TM~\cite{pc4}           & 1999       & 14         & 0.109 5(3)           & -       & -                     \\
            Duality~\cite{duity2001}       & 2001       & -         & 0.110~028          & -             & -                        \\
            TM~\cite{pc2} & 2001       & 20        & 0.109~4(2)         & 0.75(2)       & 0.18(1)                    \\
            TM~\cite{merz2002}             & 2002       & 64        & 0.109~3(2)         & $0.67(3)$ & 0.183(3)                         \\
           
            TM~\cite{TM2006}               & 2006       & 14        & 0.109~3(4)         & 0.67(2)       & 0.181(1)                  \\
            TM~\cite{TM2006_1}             & 2006       & 16        & 0.109~4(2)   & -             & 0.18(1)                       \\
             Monte Carlo (MC)~\cite{MC2008}               & 2008       & 64        & 0.109~19(7)  & 0.655(15)     & 0.180(5)                  \\
            MC~\cite{MC2009}              & 2009       & 64        & 0.109~17(3)  & 0.66(1)       & 0.177(2) \\
            TM~\cite{TM2009}              & 2009       & 14        & 0.109~35(20) & 0.64(2)       & -                      \\
            
            pTRG~\cite{pTRG}              & 2014       & 128       & 0.109~17(22) & 0.642(22)                        & -             \\
             Time-Evolving
Block Decimation (TEBD)~\cite{TEBD}              & 2020       & 300       & 0.109~96(6) & -     & -                                                                                     \\
            \hline
            {TNMC, present work}        & 2024 & 1024 & 0.109~26(2)
                                          & 0.67(1)   & 0.180(1)                   & 0.560(15)                          \\
        \end{tabular}
    \end{ruledtabular}
\end{table*}

The $\pm J$ {random bond} Ising model is a prototype of spin glasses with random couplings,
and the schematic phase diagram on 
the square lattice is depicted in Fig.~\ref{fig:phase2},
which is symmetric with respect to $p=1/2$ due to the bipartite property 
and is thus restricted to $p \leq 1/2$ only.
There is a special line with gauge symmetry~\cite{nishimori1981internal},
\begin{equation}
e^{-2/T}=p/(1-p) \; ,
\end{equation}
denoted as the dashed-dotted line in Fig.~\ref{fig:phase2}.
Along this so-called Nishimori line,  many exact results can be obtained.
Note that the {spin glass} (SG) phase transition cannot occur at any finite temperature $T$
in two dimensions, and there are only ferromagnetic (FM) and paramagnetic (PM) phases 
for finite $T$. 
At zero temperature ($T=0$), as the strength of bond disorder increases, 
the system undergoes a phase transition at $p_c\approx 0.109$ (marked as the blue circle),
entering from the FM into the SG phase~\cite{wang2003confinement, Amoruso-domain-wall}.
The line of finite-$T$ phase transitions between the FM and PM phases
intersects with the Nishimori line at a multi-critical Nishimori point (MNP),
and, accordingly, is separated into two line segments. 
The {higher} $T$ line segment starts from the pure Ising transition point~\cite{onsager}
at $p=0$ and $T_c=2/\ln(1+\sqrt{2})$ (marked by a star symbol labeled as PF Ising), 
and, as $p$ increases, the critical temperature $T_c$ decreases till MNP.
The phase transition along this {higher} $T$ line segment 
belongs to the weak-disordered universality class:  
disorder gives only rise to logarithmic corrections 
to the random diluted Ising  (RDI) critical behavior~\cite{RDI}.
Along the lower-$T$ line segment of transition, 
as $T$ decreases, the critical value $p_c$ slightly decreases 
and ends at the zero-temperature transition point between the FM and the SG phase. 
The phase transition for $T<T_{\text{MNP}}$ 
belongs to the strong-disorder Ising (SDI) universality class.

Numerous simulations have been conducted to explore the $\pm J$ 
{random bond} Ising model in 2D. 
For this problem, the rugged landscape results in a slow mixing of the Markov chain with local updates.
Nevertheless, for the {spin glass} case with $p=0.5$, a clever geometric cluster method,
making use of the paramagnetic state, 
has been developed to interchange spin states between different replicas of the Ising 
configurations~\cite{houdayer_cluster_2001,zhu_efficient_2015}.
Together with the parallel tempering technique, 
the simulations have been performed up to systems of size $100 \times 100 $ 
down to temperature $T=0.1$. {Furthermore}, an extension of the Frank-Lobb bond-propagation algorithm can simulate the 2D random bond Ising up to the size $128\times 128$~\cite{loh_efficient_2006}.
Extensive MCMC simulations have also been carried out along the Nishimori line 
to locate the MNP, for which the geometric cluster method 
becomes significantly less effective. 

The results from the year 1987~\cite{tm1987,mcrg} to the present (year 2024) 
about the critical properties of the MNP are summarized in \autoref{tablelist} from Refs.~\cite{Ueno1991,pc5,pc4,pc2,duity2001,merz2002,TM2006_1,TM2006,MC2008,TM2009,MC2009,pTRG,tm1987,TEBD,mcrg}.
The early estimate of the MNP is about $p_c=0.11(1)$~\cite{mcrg}.
In Refs.~\cite{MC2008,MC2009}, a {large scale} MC simulations 
with the parallel tempering technique
was performed along the Nishimori line: 
the largest system size was $L=64$, about $10^6$ realizations of bond disorders were generated 
for each point of parameters $(T,p,L)$, 
and, for each disorder realization, about $10^6$ samples were taken. 
The estimate of the MNP 
was improved to be $p_c=0.10919(7)$~\cite{MC2008} and $0.10917(3)$~\cite{MC2009}. 
It is emphasized that, despite such extensive studies, 
the physics of the random bond Ising model is far from being well understood. 
At the MNP, since it is a multi-critical point, one would expect that 
there are additional relevant thermal and magnetic renormalization exponents 
beside $y_1$ and $2-\eta$ in \autoref{tablelist}. 
It remains also open how to characterize the critical 
behaviors of the random bond Ising model in the framework of conformal field theory.
To make further progress, efficient numerical methods can play an important role.

In this work, we demonstrate that the TNMC method is efficient and effective in studying disordered systems.
We show that with a relatively small bond dimension with $D=16$, 
the TNMC simulation at the MNP has an acceptance probability $p_a \approx 0.67$ 
for $L=512$ and $p_a \approx 0.21$ for $L=1024$, apparently 
no parallel tempering is needed in the TNMC simulations. 
Since the configurations proposed by tensor network contraction and sampling are independent, the corresponding autocorrelation time is only 3--5 sweeps, and critical slowing down is nearly absent even for large systems.
Similar to the Metropolis-Hastings algorithm, the TNMC method also exhibits O$(N)$ 
computational complexity per sweep ($N=L^2$ is the system volume), 
but it has a significantly larger prefactor $\propto D^2$. 
Taking into account the exponential slowing down with respect to system volume and inverse temperature for the Metropolis-Hastings algorithm, we find the constant cost—roughly 1000 times for $D=16$—to be acceptable, particularly for large system sizes and low temperatures.
The parameters of our simulations are listed in Table~\ref{tableparameter}.
For $L=512$ and $1024$, the number of disorder realizations is only $5000$ and $500$, 
respectively, and, as a consequence, the statistical errors 
of physical observables are still relatively big.  
The MNP determined in our simulations is $p_c=0.10926(2)$, 
which is slightly larger than $p_c=0.10919(7)$ in Ref.~\cite{MC2008} and $p_c=0.10917(3)$ in Ref.~\cite{MC2009}.
Although our MC data for $L=96$, $L=128$, and $L=256$ gives a hint that 
the finite-size extrapolation might be slightly underestimated in Refs.~\cite{MC2008,MC2009},
a conclusive statement would request a significantly larger number 
of disorder realizations. Achieving this would necessitate access to a larger-scale computing facility, beyond our current small cluster, and we defer this investigation to future work.

\subsection{Acceptance ratio  of TNMC}
\begin{figure}[h]
\includegraphics[width =1.1\linewidth]{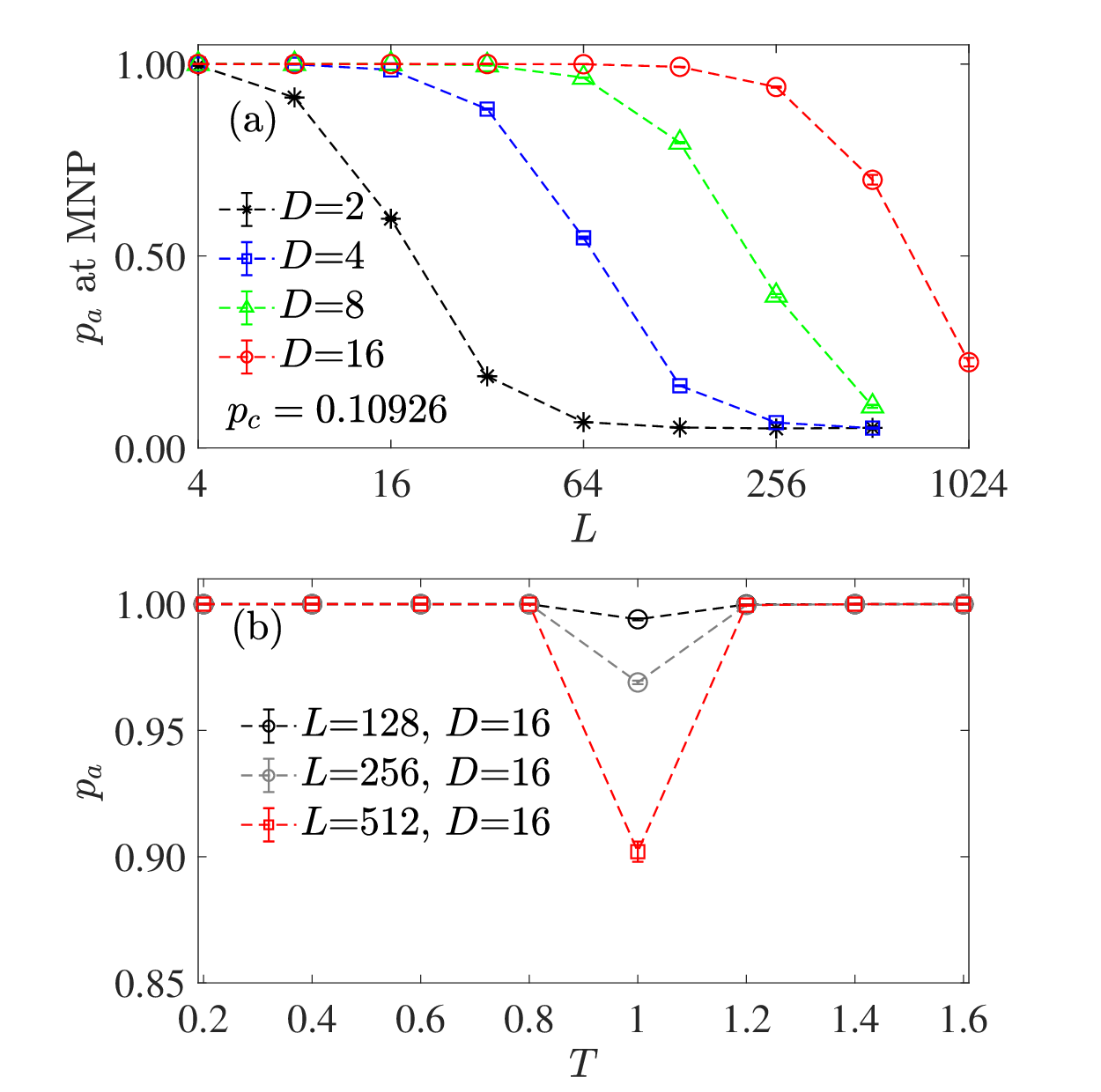}
    \caption{ The acceptance probability $p_a$ for tensor network proposals with various bond dimensions and lattice sizes (a) and for tensor network proposals with various temperatures for the fixed bond dimension $D=16$ (b).} 
    \label{fig:pa}

\end{figure}

We first study the acceptance probability $p_a$ of the Metropolis scheme 
in the TNMC simulation with system size from $L=8$ to $1024$. 
Since the configuration proposals from tensor network computation are independent, the autocorrelation time is approximately equal to the inverse acceptance probability $1/p_a$.
The results are shown in Fig.~\ref{fig:pa} (a) for the MNP ($T$=0.9531, $p=0.10926$). 
As expected, $p_a$ is quickly enhanced as the bond dimension $D$ increases, 
since a larger $D$ means a more accurate computation of the conditional probabilities. 
Remarkably, with $D=16$, $p_a$ remains at approximately $100\%$ for sizes up to $L=128$, 
and $p_a$ exceeds $75\%$ for $L=512$. 
Even for $L=1024$, the acceptance probability is still about $25\%$, indicating a very small autocorrelation time $\tau \approx 4$. 
In contrast, in the Metropolis-Hasting simulation of a disordered system, 
the autocorrelation time is expected to exponentially increase as $\tau \sim e^{\# N/T}$, 
where $N=L^d$ is for system volume and $\#$ represents a non-zero constant. 
Fig.~\ref{fig:pa}(b) shows the results of $p_a$ along the Nishimori line, with $D=16$ being fixed. 
We can see that $p_a$ has a minimal value around the critical point $T_c=0.9531$, 
which decreases as $L$ increases. This critical slowing down is due to the divergence of correlation length $\xi$. 
In the language of quantum field theory, critical systems are {\it gapless}: 
the gap between the ground and the excited states vanishes as $L$ increases ($\xi$ is just the inverse of the gap). 
Nevertheless, as shown in Fig.~\ref{fig:pa} (a,b), the $p_a$ values in practical calculations are still significantly large even for $L=1024$, 
implying that the critical slowing down is nearly absent. 
At low temperature, there are two global free-energy minima related by spin-flipping symmetries,
which makes the entire temperature region (in the absence of an external field) 
effectively a line of first-order phase transitions. 
The high acceptance ratio suggests that the TNMC method might be a powerful tool for 
exploring first-order phase transitions.

\begin{figure}[h]
\includegraphics[width =1.1\linewidth]{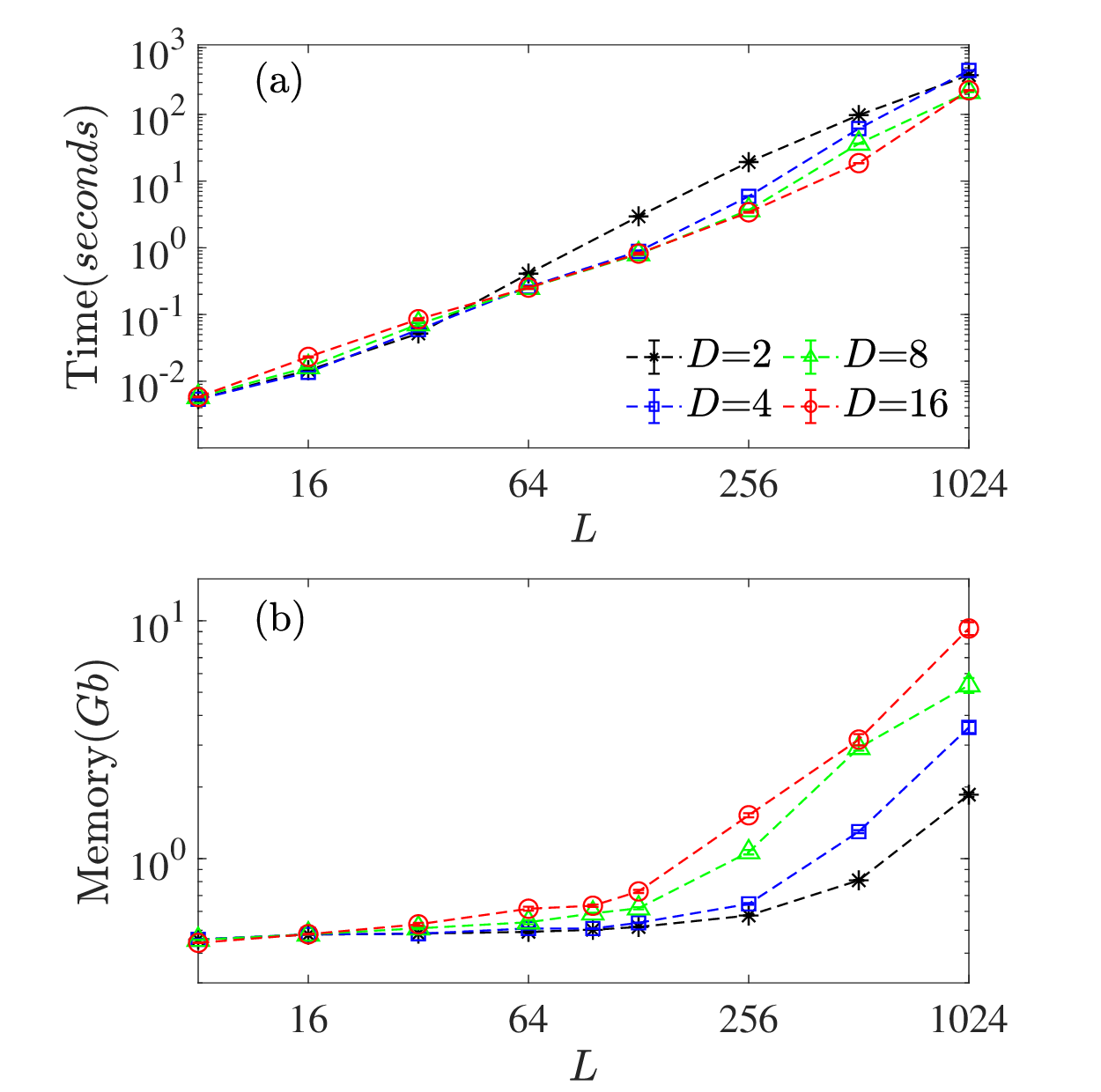}
    \caption{ CPU time (a) and memory usage (b) 
     at the MNP  for various lattice sizes $L$ and bond dimensions $D$. }
\label{fig:time_memo}
\end{figure}

In the TNMC, each sampling of a local spin involves the product computation of a few $D \times D$ matrices, 
and, thus, generating a proposed configuration would cost CPU time $ t \propto D^2 N$. 
As a consequence, albeit still being {of order} $O(N)$, the actual CPU time per sweep in TNMC 
can cost more CPU time than in the Metropolis-Hasting algorithm, by a factor of a few hundred. 
This compensates for the enhancement of the autocorrelation time gained by the TNMC. 
In Fig.~\ref{fig:time_memo} (a), for each effectively independent sample, 
the time costs are evaluated at the MNP for various lattice sizes $L$ and bond dimensions $D$.
Computation time is defined by the actual simulation time per sample (on an Intel CPU, 2.6GHz) 
divided by the acceptance probability of that sample, i.e., 
 \begin{equation}
     \text{computation time} =\frac{
\text{actual simulation time}}{
\text{acceptance probability}}.
 \end{equation}
The computation time for each independent sample increases with size, 
but in the double logarithmic plot, the computation time appears to be slightly sub-linear.  
When the size $L$ exceeds 32, the computation time for obtaining valid samples with a larger bond dimension $D=16$ is less than that for $D=2$. 
Although the actual simulation time for $D=16$ is longer, the relatively high acceptance probability results in a shorter valid computation time, 
suggesting that a fine-tuning of $D$ is needed for an optimized simulation efficiency. 
For the size $L=1024$, the computation time for one valid sample is 226 seconds for a single thread, approximately 3.8 minutes. 
{Computing} 10,000 valid samples would require approximately 26.5 days.
 
To further illustrate the practicality of the TNMC, the required computer memory is shown in Fig.~\ref{fig:time_memo} (b). 
When $D=16$, the computer memory for sizes $L=256$, 512, and 1024 is about 1.5, 3.1, and 9.8 Gigabytes (Gb), respectively. 

\subsection{Sampled quantities}

To determine the critical point and the critical exponents, we measure the following quantities.

\begin{itemize} 
    \item The overlap order parameter
    \begin{equation}
        q = \frac{1}{L^2} \sum_{i} s_i^{(1)} s_i^{(2)},
    \end{equation}
    where the superscript 1 and 2 mark two different replicas.
    \item The Binder ratio
    \begin{align}
        Q_q &=  \left[\expval{q^4}\right]/\left[\expval{q^2}^2\right],\label{eq:binderEA}
    \end{align}
    where $\left[~ \right]$ represents the averages over various disorder realizations 
    {and} $\langle~ \rangle$ is for ensemble average.
    \item The magnetic susceptibility for overlap order parameter $q$,
    \begin{align}
       \chi_{ q} &= \frac{L^2}{T} \left(\left[\expval{q^2}\right] - \left[\expval{q}\right]^2\right) . \label{eq:susEA}
    \end{align}
\end{itemize}

Near the critical point, these physical quantities asymptotically satisfy the following finite-size scaling (FSS) functions
\begin{align}
    Q_q &= \mathcal{F}_Q\left[L^{1/\nu}(p-p_c)\right],\label{eq:fssQq} \\
    \chi_q &=  L^{2-\eta} \mathcal{F}_\chi\left[L^{1/\nu}(p-p_c)\right],
\end{align}
where $1/\nu=y_{1}$ is the thermal renormalization exponent and $\eta$ is the magnetic anomalous dimension,
$\mathcal{F}_Q$ and $\mathcal{F}_\chi$ are the analytic scaling functions
that are expected to be universal. 
In practice, there also exist additional finite-size corrections that should be taken into account in the least-squared criterion fitting of the TNMC data. 

Since the TNMC simulation is carried out {on} a $L \times L$ square lattice with free boundary conditions, 
we also study the surface (edge) critical behaviors.
We measure the surface (edge) magnetization along the four free boundaries as:
\begin{equation}
    m_{\text {edge }}=\frac{1}{2 L} \sum_{i=\frac{L}{4}}^{\frac{3 L}{4}}\left(s_{i, 1}+
    s_{1, i}+s_{L, i}+s_{i, L}\right) \;, 
\end{equation}
where, to suppress the influence of spins on the corners, four line segments of length $L/2$ 
are selected around the center of each edge for the measurement. 
Then, the surface magnetic susceptibility is defined as $\chi_\text{edge} = L \left[ \langle  m_\text{edge}^2 \rangle \right]$, 
and is expected to scale as $\chi_\text{edge} \sim L^{1-\eta_\text{edge}}$ at the critical point with $\eta_\text{edge}$ the edge magnetic anomalous dimension. 

\subsection{Numerical Results}

 To determine the MNP point, we simulate the {random bond} Ising model around the MNP along the Nishmori line.  
Simulations are performed for system sizes from $L=8$ to 1024.
For each $L$ with $L \leq 256$, a number of probabilities $p$ are chosen, and a large number of disorder realizations are taken for each probability. 
Considering that our computational resources are rather limited, 
the simulations for $L=512$ and 1024 are less extensive and mostly carried out at 
the estimated critical point $p_c$. 
While these {large} $L$ data do not significantly improve the precision of $p_c$, 
they are important in estimating the bulk and the edge magnetic anomalous dimensions. 

Then, at $p_c$,  we perform the least-squares criterion fits for the bulk and the edge magnetic susceptibility, $\chi_q$ and $\chi_{\text{edge}}$, 
and obtain the bulk and edge magnetic anomalous dimensions $\eta$ and $\eta_{\text{edge}}$. 
To our knowledge, $\eta_{\text{edge}}$ is obtained for the first time, 
where the data of $L=512$ and $1024$ play an useful and necessary role. 

Finite-size scaling (FSS) analysis is used to determine the critical point $p_c$ and the exponents, 
where finite-size correction terms are also taken into account. 
During the FSS analysis, generally, the preferred fitting procedure is as follows. 
Starting with the smallest $ L_{\text{min}}$, the fit should be deemed reasonable if further increases in $L_{\text{min}}$ 
do not lead to a substantial reduction in the {$\chi^2$} value by more than one unit per degree of freedom. 
Practically, ``reasonable" implies that {$\chi^2/\text{DF} \approx 1$}, where DF represents the number of degrees of freedom. 

The sample-to-sample fluctuations and the issue of self-averaging 
for the random bond and random field Ising models
were discussed in Refs.~\cite{fytas2008selfaveragingcriticalitycomparativestudy,Fytas2011}.
Along the Nishimori line, self-averaging is expected to hold  
since the system is a ferromagnet at low $T$.
Nevertheless, it might be interesting in future to study how the probability distribution of 
the order parameter evolves as a function of temperature and system size.


\subsubsection{The multi-critical Nishimori point}

As listed in \autoref{tablelist} from Refs.~\cite{Ueno1991,pc5,pc4,pc2,duity2001,merz2002,TM2006_1,TM2006,MC2008,TM2009,MC2009,pTRG,tm1987,TEBD,mcrg}, 
the pursuit for high-precision critical value $p_c$  for the MNP has a long history. 
Before 1996, the error bars for $p_c$ were typically reported to the second or third decimal place~\cite{mcrg, tm1987, Ueno1991, pc5}. 
Between 1999 and 2006 \cite{pc4, pc2, duity2001}, they were refined to the fourth decimal place. 
Starting from 2008, advancements in the modern computer industry and skillful methods for data analysis
have led to further precision, with error bars reaching the fifth decimal place~\cite{MC2008, MC2009}. 
However, for the transfer matrix (TM) method, constrained by a width limitation of $L\le 14$, the error bars remained at the fourth decimal place \cite{TM2009}.

In the tensor network methods, specifically the topological invariant tensor network renormalization (TRG) method \cite{pTRG}, 
system sizes of up to $128\times 128$ have been achieved, yet the error bars remain at the fourth decimal place. 
In 2020, Ref.~\cite{TEBD} employed the time-evolving block decimation (TEBD) method, 
but the result $p_c=0.109~96(6)$ is inconsistent with the best MC estimate $p_c=0.109~17(3)$~\cite{MC2009}, 
if the quoted error margins are taken seriously. 

\begin{table}[ht]
  \caption{For each system size $L$ we equilibrate $N_{\text{therm}}$ steps and then measure for at least $N_{\text{run}}$ configurations. $N_{\text{s}}$ is the number of disorder realizations and $D$ is the cutoff bond dimension of the matrix product states.}
  \tabcolsep=0.5 cm
  \begin{tabular}{lllll}
  \hline
  \hline
  $L$ & $D$ & $N_{\text{s}}/10^3$ & $N_{\text{run}}$ & $N_{\text{therm}}$\\
  \hline
  8   &   8   &   1000     &    10      &   0 \\
  12  &   8   &   1000     &    10      &   0 \\
  16  &   8   &   1000     &    10      &   0 \\
  24  &   8   &   1000     &    10      &   0 \\
  32  &   8   &   1000     &    10      &   0 \\
  64  &   8   &   1000     &    20      &   10 \\
  96  &   8   &   200      &    30      &   20 \\
  128 &   16  &   20       &    100     &   0 \\
  256 &   16  &   15       &    100     &   50 \\
  512 &   16  &   5        &    150     &   80 \\
  1024&   16  &   0.5      &    300     &   100 \\
  \hline  

  \hline
  \hline
  \end{tabular}
  \label{tableparameter}
\end{table}

\begin{table*}[ht]
    \caption{Fitting results for the Binder ratio $Q_q$ using the ansatz Eq.~\eqref{eq:fssQq}.}
    \tabcolsep=0.2 cm
    \begin{tabular}{ccllllllllll}
    \hline
    \hline
    Obs. & $L_{\text{min}}$ & $\chi^2$/DF & $p_c$ & $y_1$  & $Q_0$ & $a_1$ & $a_2$ & $b_1$ & $b_2$ & $c_1$ & $\omega$  \\
    \hline
    $Q_{q}$  
    & 8  & 70.5/99 & 0.109~224(8)  & 0.663(5) & 1.447~9(3) & -2.66(5) & 5.2(9) & -0.27(1) & 0.33(7) & 0 & 1  \\
    & 12  & 59.1/88 & 0.109~26(1)  & 0.666(5) & 1.449~0(5) & -2.65(6) & 5.1(8) & -0.34(2) & 0.9(2)  & 0 & 1  \\
  
    & 8  & 70.3/98 & 0.109~223(8)  & 0.67(1) & 1.447~9(3) & -2.6(1) & 4.9(9) & -0.27(2) & 0.35(7) & -1.0(9) & 1  \\
    & 12 & 58.9/87 & 0.109~26(1)  & 0.67(1) & 1.450~6(6) & -2.7(2) & 5(1) & -0.35(2 ) & 0.9(2) & -1(1) & 1  \\
  
    \hline
    \hline
    \end{tabular}
    \label{tableQq}
\end{table*}

The details of our TNMC simulations are summarized in Table~\ref{tableparameter}. 
The system sizes are taken as $L=$8, 12, 16, 24, 32, 64, 96, 128, and 256 in the range $p \in [0.1080,0.1100]$ along the Nishimori line, and the simulations for $L=$512 and 1024 are only at the estimated MNP $p_c=0.109 26$. The number of disorder realizations for each data point is $10^6$ for $L \leq 64$, and decreases rapidly for $L \geq 96$ due to our limited computer resources.

\begin{figure}[t]
\includegraphics[width=1.0\linewidth]{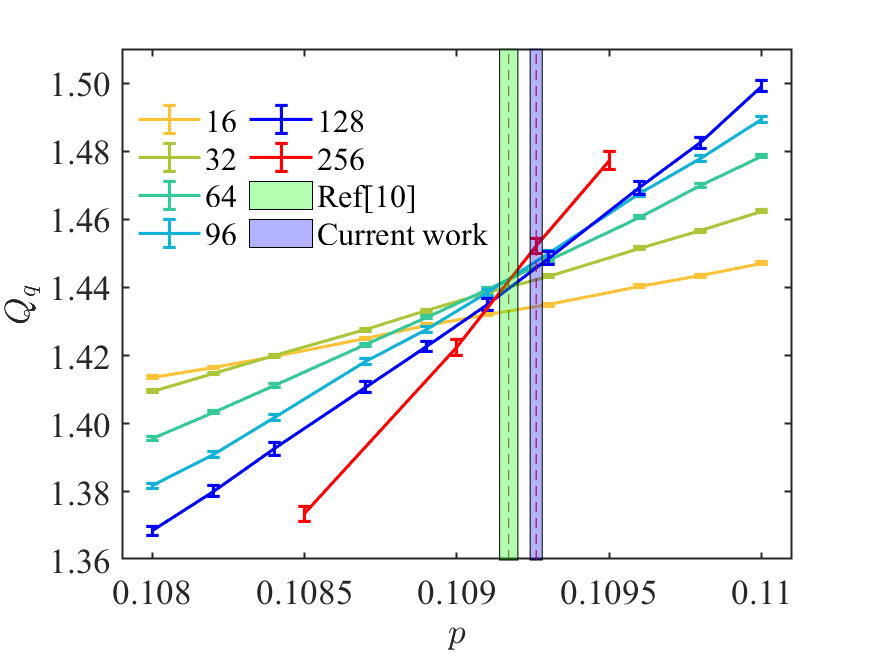}
\caption{Finite-size scaling of the Binder cumulant $Q_q$ along the Nishimori line. 
The light blue shaded area indicates the fitting result of $p_c=0.10926(2)$ obtained in this article, the critical exponent is obtained as $y_1=0.67(1)$. The green shaded area represents the fitting result $p_c=0.10917(3)$ in Ref.~\cite{MC2009} for comparison.} 
\label{fig:2DRBIMQqNishimori}
\end{figure}
    
The $Q_q$ data for $L \leq 256$ are shown in Fig.~\ref{fig:2DRBIMQqNishimori}. As $L$ increases, the intersection points move toward a larger value of $p$, clearly indicating $p_c > 0.1090$. Note that, the intersection point between the $L=98$ and 128 data lines seems to be $p \approx 0.1093$, it does not give much information due to the significant statistical error margins in the $L=256$ data. According to the finite-size scaling form in Eq.~(\ref{eq:fssQq}), we perform least-square fits to the $Q_q$ data by 
\begin{align}
    Q_q = &Q_0 + a_1(p-p_c)L^{y_1} + a_2(p-p_c)^2L^{2y_1} \nonumber\\
            &+ b_1L^{-\omega} + b_2L^{-2\omega} + c_1(p-p_c)L^{y_1-\omega},
    \label{ansatzQq}
\end{align}
where $Q_0$ is a universal value and $a_1$ and $a_2$ are nonuniversal constants. 
The term $b_1$ accounts for the finite-size corrections with exponent $\omega>0$,  $b_2$ accounts for the second corrections with exponent $2\omega$. 
The $c_1$ term accounts for the crossing effect between finite-size corrections 
and the scaling variable $(p-p_c)L^{y_1}$.

In Ref. \cite{MC2008}, the correction-to-scaling exponent was not well-established, 
so the authors treated it as a free parameter. 
By fitting various quantities, they concluded that $\omega \gtrsim 1$. 
However, it was difficult to obtain a definite value of $\omega$ that converges as 
$L_{\text{min}}$ increased. 

In the current work, despite that the existence of finite-size corrections 
is displayed in Fig.~\ref{fig:2DRBIMQqNishimori}, 
the MC data are not sufficiently accurate to give a reliable estimate of 
the correction exponent $\omega$. 
Meanwhile, the free boundary conditions used in this study may lead to 
different correction behavior compared to the results obtained 
with periodic boundary conditions~\cite{MC2008}.

Therefore, we simply take $\omega=1$ as an assumption and fix this value,
as partly supported by the bulk susceptibility in Fig.~\ref{fig:chi}.
The fitting results are shown in Table~\ref{tableQq};
taking other values in range $\omega \in [1,2]$ has only small effects.
When $L_{\rm min}$ is increased from 8 to 12, the estimated critical point increases from $p_c=0.109\,224(8)$ 
to $0.109\,26(1)$; since the residual $\chi^2$ per degrees of freedom drops significantly, 
the finally quoted value of $p_c$ is taken as $0.109\,26$. 
In addition, taking into account that additional corrections might not be included in the fitting formula, 
we double the statistical fitting error in the final estimate $p_c=0.109\,26(2)$.
The fitting result for $y_1$ is rather stable when $L_{\rm min}$ is increased. 
For further testing, we also try to include the cross term with coefficient $c_1$, 
which is found to play a negligible role. 
Thus, our final results are taken as $p_c=0.109\, 26(2)$ and $y_1=0.67(1)$, 
which are more or less consistent with the most recent MC estimates in Table~\ref{tablelist}.

\subsubsection{Bulk magnetic anomalous dimension $\eta$}

To estimate the bulk magnetic anomalous dimension $\eta$, we fit the data of the critical magnetic susceptibility $\chi_q$, collected at the MNP $p_c =0.109\,26$, to 
\begin{equation}
    \chi_q= L^{2-\eta}(a_0+b_1L^{-\omega}) \; ,
    \label{eq:chi}
\end{equation}
where $\omega=1$ is fixed. The results of the fits are shown in Table~\ref{tablechi}, which gives $\eta = 0.180(1)$.

\begin{table}[ht]
    \caption{Estimating $\eta$  from  the data
    $\chi_q$ at MNP.}
    \tabcolsep=0.15 cm
    \begin{tabular}{cllllll}
    \hline
    \hline
     $L_{\text{min}}$& $\chi^2$/DF & $a_0$ & $b_1$ & $\eta$ & $\omega$\\
    \hline  
    16 & 1.9/7 & 0.097~8(7) & 0.065(7) & 0.180(1) & 1  \\
    24 & 1.7/6 & 0.098(1) & 0.05(1) & 0.180(1) &  1  \\
    \hline
    \hline
    \end{tabular}
    \label{tablechi}
\end{table}

\begin{figure}[h]
\includegraphics[width=1.0\linewidth]{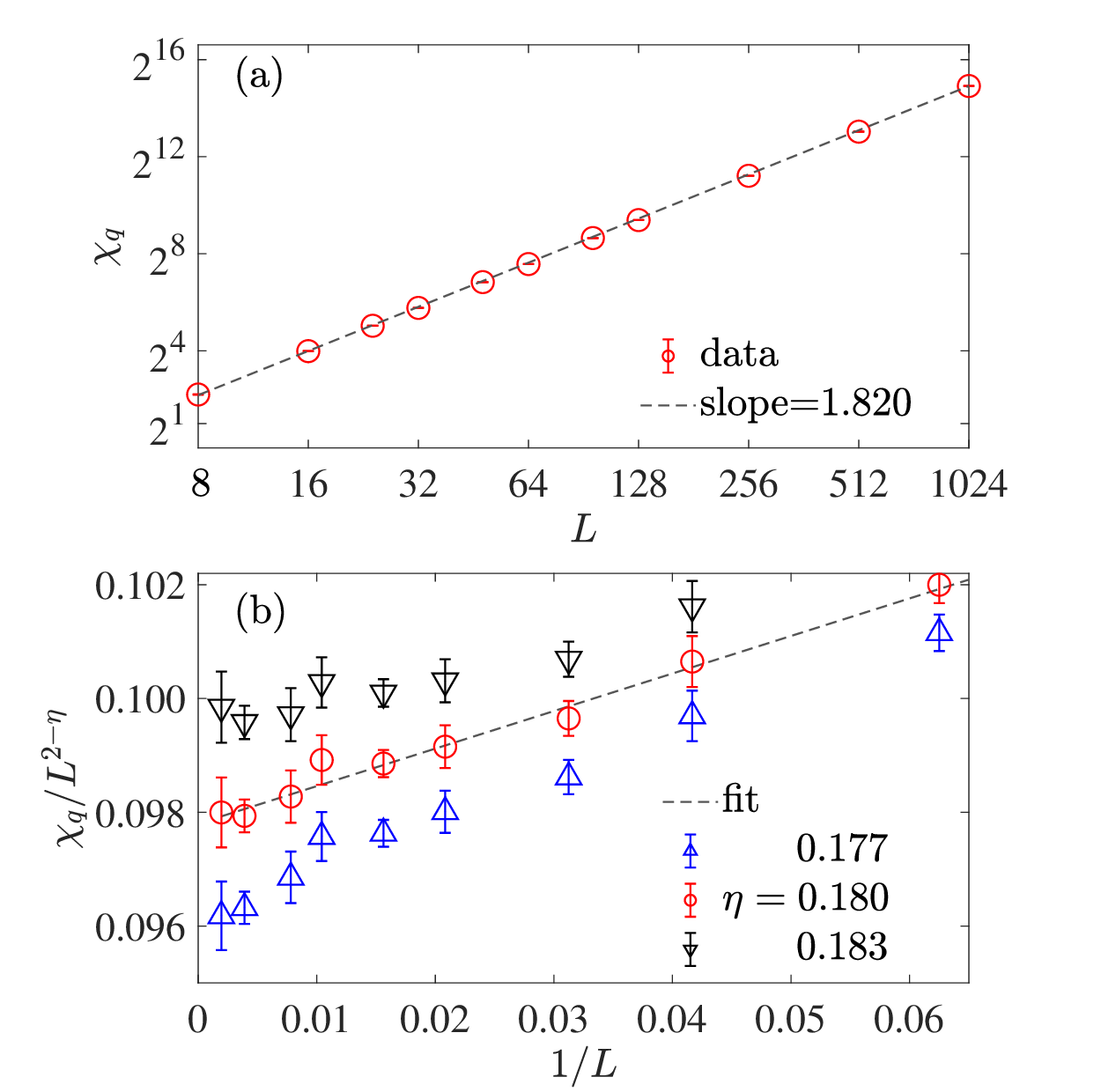}
 \caption{ Scaling behavior of $\chi_q$ and bulk magnetic exponent $\eta$ at $p_c$. (a) According to Eq.~(\ref{eq:chi}), the plot of  $\chi_q$ versus $L$ is depicted on a log-log scale, revealing a linear relationship between them. (b) $\chi_q/L^{2-\eta}$ versus $1/L$ is presented, and the best outcome indicates $L^{-\omega} = 1/L$. The other two bending data indicate the reliability of the center value $0.180$ and the error bar $0.001$ quoted in the final analysis.}
    \label{fig:chi}
\end{figure}

Figures~\ref{fig:chi} (a) displays the $\chi_q$ data at $p_c$ versus $L$ on a log-log scale. 
The collapse of the MC data, from $L=8$ to $1024$, onto the straight line suggests that 
finite-size corrections are not severe and $\eta=0.180$ is a good estimate. 
This is also reflected by Table~\ref{tablechi}, where the correction amplitude $b_1$ 
is rather small. 
To further check the reliability of the fitting result, in Fig.~\ref{fig:chi}(b) 
we plot the rescaled susceptibility $\chi_q/L^{2-\eta}$ versus $1/L$, 
where the values of $\eta$ are taken as $0.180$ and $0.180 \pm 3 \sigma$ 
with the error bar $\sigma = 0.001$ from the fits. 
The approximately straight line for $\eta=0.180$ suggests that the leading finite-size correction exponent 
is indeed about $\omega=1$.  
Further, the downward and the upward bending tendencies for large system sizes, 
respectively for $\eta=0.177$ and $0.183$, indicate that the error bar $\sigma=0.001$ 
is more or less reliable. On this basis, we take the finally quoted estimate as $\eta = 0.180 (1)$, consistent with the latest MC result $\eta=0.177(2)$~\cite{MC2009}.

\subsubsection{Surface magnetic anomalous dimension $\eta_{\text{edge}}$}

Surface critical phenomena have long been a focal research subject in statistical physics~\cite{surface_binder,landau_binder}. 
While extensive studies have been carried out in various systems with pure interactions~\cite{3d_ising_conformal,on,tri_potts,o4,3dxy,potts,edge_blume_capel,sublattice}, 
research attention in disordered systems, particularly at the MNP, remains scarce. 
In the current TNMC simulations, since free boundary conditions are applied, the edge magnetic susceptibility $\chi_{\text{edge}}$ is then measured without additional effort. 

\begin{figure}[h]
\includegraphics[width=1.0\linewidth]{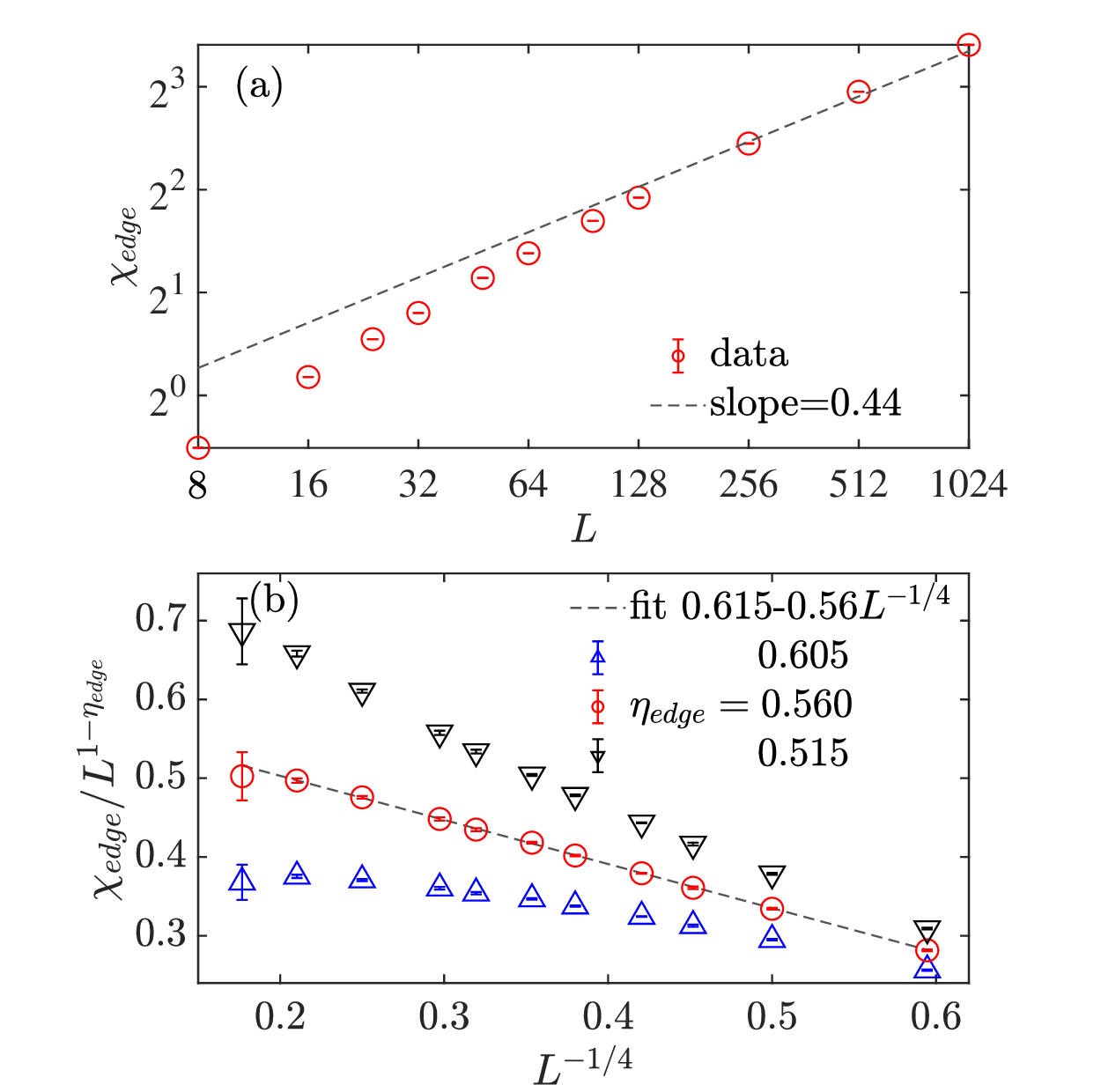}
\caption{Scaling behavior of $\chi_\text{edge}$ and edge magnetic exponent 
$\eta_{\text{edge}}$ at $p_c$. (a)   $\chi_{\text{edge}}$ vs $L$, (b) fitting of Eq.~(\ref{eq:chi-edge}), yielding   $\eta_{\text {edge }}=0.560(15)$. The other two bending data indicate the reliability of the center value $0.560$ and error bar $0.015$  quoted in the final analysis.}
\label{fig:chi-edge}
\end{figure}

The log-log plot of the $\chi_{\text{edge}}$ data versus $L$ is shown in Fig.~\ref{fig:chi-edge}(a), 
where the dashed line with slope $0.44$ is from the fit. 
For small system sizes, the MC data are clearly away from the straight line, indicating that finite-size corrections are severe and the {large} $L$ data play an important role in determining the edge magnetic anomalous dimension $\eta_{\text{edge }}$.

The $\chi_{\text{edge}}$ data are fitted to
\begin{equation}
\chi_{\text {edge }}=L^{1-\eta_{\text {edge }}}\left(a_{0}+b_{1} L^{-\omega}\right) \; .
\label{eq:chi-edge}
\end{equation}
Since there is no established expression for the correction exponent $\omega$,
we tested various fitting approaches, as shown in Table~\ref{tab:edge}. 
Initially, we attempted to fit the data using Eq.~(\ref{eq:chi-edge}) with all parameters free. 
As we increased the minimum system size $L_{\text{min}}$, the $\chi^2/\text{DF}$ remained close to 1. 
However, the error in the correction exponent $\omega$ was substantial, 
preventing us from determining a precise value for $\omega$. 
Despite this, the results suggest that the leading correction exponent $\omega$ is approximately 
within the range $\omega \in (0.2, 0.4)$, significantly larger than the bulk susceptibility value of $\omega = 1$. 

\begin{table}[ht]
    \caption{The fitting details for $\chi_{\text {edge }}$ in Eq.~(\ref{eq:chi-edge}).}
    \tabcolsep=0.15 cm
    \begin{tabular}{cllllll}
    \hline
    \hline
        $L_{min}$ & $\chi^2$/DF & $a_0$ & $b_1$ & $\eta_{\text{edge}}$ & $\omega$\\
    \hline
        8   & 7.1/7  & 0.5(2) & -0.5(1) & 0.55(3) & 0.28(14)  \\
        16  & 6.7/6  & 0.5(2) & -0.47(17) & 0.54(4) & 0.33(25)  \\
        8   & 7.1/8  & 0.537(7) & -0.49(1) & 0.548(2) & 0.3  \\
        16  & 6.8/7  & 0.54(1) & -0.50(2)  & 0.550(4) & 0.3  \\
        8   & 7.0/8  & 0.615(9) & -0.56(1) & 0.560(2) & 0.25  \\
        16  & 6.9/7  & 0.62(1)  & -0.57(3) & 0.561(4) & 0.25  \\
    \hline
    \hline
    \end{tabular}
    \label{tab:edge}
\end{table}

Subsequently, we fixed $\omega$ at $0.3$, resulting in $\eta_{\text{edge}} = 0.550(4)$. We also tried fixing $\omega$ at $0.25$, yielding $\eta_{\text{edge}} = 0.561(4)$. Nevertheless, taking into account the uncertainty arising from unknown correction terms, we take the final estimate to be $\eta_{\text{edge}} =0.560 (15)$. 

The rescaled edge susceptibility $\chi_{\text {edge }}/L^{1-\eta_{\text {edge }}}$ 
is plotted versus $L^{-1/4}$ in Fig.~\ref{fig:chi-edge} (b), 
where the $\eta_{\text {edge }}$ value is taken as $0.515$, $0.560$, and $0.605$. 
For $\eta_{\text {edge }} = 0.515$ and $0.605$, the deviation from the approximately straight line 
indicates the reasonableness of the quoted {error} bar in $\eta_{\text {edge }} =0.560(15)$.

\section{discussion}
\label{sec:conclusion}

In conclusion, we have investigated the random bond Ising model in two dimensions using the TNMC algorithm.
We have obtained comparable or better precise estimates of the NMP and critical exponents of the model, surpassing previous results with much larger system sizes. In particular, we show that the TNMC simulation can be performed for system sizes up to $L=1024$, which is significantly greater than the system size $L=64$ in the previous studies. Our results $p_c=0.109\,26(2)$ and $\eta=0.180(1)$ also suggest that systematic errors are slightly underestimated 
in the finally quoted values $p_c=0.109\,17(3)$ and $\eta=0.177(2)$ in the previous MCMC studies~\cite{MC2009}. 
We observe that in the large system simulations of the random bond Ising model using TNMC, the critical slowing down is nearly absent, giving a fixed autocorrelation time, 
and the requested computer memory is easily affordable within modern computers. 

Taking into account that the TNMC method is recently developed 
and it is hybrid algorithm, we present a detailed description
and illustrate some basic concepts using the one-dimensional Ising model in the Appendix. 
Our Julia code is available on Github with a comprehensive tutorial~\cite{github}. 
We feel it valuable for readers who are interested in further applications 
and developments of the TNMC method. 
Moreover, we provide a systematic study of the performance of the TNMC method as well as its cost in computing time and computer memory. 

Compared with the conventional MCMC methods, the computational complexity for generating a sample in TNMC is much larger. The CPU time per sweep in TNMC is in general $ t \approx A D^2 N$, where $D$ is the maximal bond dimension of the tensor network, $N$ is the system volume and $A$ is a constant in our simulations of 2D RBIM.  which underscores the importance of carefully choosing the number of samples for each disorder realization. Since the CPU time scales as $D^2$, the efficiency of TNMC simulations can be further optimized by fine-tuning the bond dimension $D$ and the acceptance probability $p_a$. Our data suggest that for $D=16$, the actual CPU time per sweep is about 1000 times larger than that for the conventional MC algorithm. As a consequence, for the systems where efficient MC update scheme exists--such as the cluster or worm algorithms--are available, the TNMC is probably not an efficient choice. However, for disordered systems with rugged energy landscapes, or systems undergoing first-order phase transitions, where the conventional MC methods suffer from exponentially growing autocorrelation time, the TNMC can provide a very useful research tool. 

The conventional tensor network methods usually demand a large truncation dimension, 
particularly {for} three-dimensional lattices at critical points~\cite{trg_3d}, and demand a large memory.
In contrast, the TNMC releases the demands for large bond dimensions, leaving the accuracy requirements to the MC sampling process.
In addition, the TNMC method can sample any physical quantities defined on the basis of configurations, 
which are useful in exploring system properties in general~\cite{binder,trg_binder}. 
Moreover, for systems that undergo continuous phase transitions and thus have divergent correlation lengths, 
it is normally challenging to extrapolate to the $D \to \infty$ limit by the traditional TN methods; 
this is probably the reason why the TEBD estimate of the MNP~\cite{TEBD}, 
$p_c=0.109 \, 96(6)$, is not fully consistent with the MC results. 
In contrast, the TNMC simulation is numerically exact, and the critical behaviors can be 
analyzed by the standard finite-size scaling theory. 
Additionally, although this work focuses on models with open boundary conditions, 
the TNMC method can be straightforwardly generalized to periodic boundary conditions.
In each TNMC step, a random pair of horizontal and vertical lines of spins is chosen,
and, by keeping these spins unchanged during the current step, 
they act as external fields to those adjacent to the chosen pair of lines. 
Finally, the remaining region, excluding this pair of lines, can be treated 
to have open boundaries. 
While this may introduce somewhat stronger correlations between samples, 
we expect that it remains acceptable for the TNMC method.

As a conclusion, we have provided a concrete example that the TNMC can serve as a powerful research tool to study disordered systems with rugged free energy landscapes. We expect the TNMC method and its extensions would find broad applications in two-dimensional systems with disorders or with first-order phase transitions. It would be more important to generalize to high-dimensional systems and quantum models, which are our current ongoing research activities. 

\vskip 1cm
{\it  Acknowledgments}  This work was supported by the National Natural Science Foundation of China (under Grant No.~12275263, 11747601, 11975294), the Innovation Program for Quantum Science and Technology (under Grant No.~2021ZD0301900), and 
the Hefei National Research Center for Physical Sciences at the Microscale (KF2021002). 
YD acknowledges the Natural Science Foundation of Fujian province of China (under Grant No.~2023J02032).
WZ acknowledges Shanxi Province Science Foundation (Grants No:~202303021221029).
PZ acknowledges Projects No. 12325501, No. 12047503, and No. 12247104 of the National Natural Science Foundation of China, and Project No. ZDRW-XX-2022-3-02 of Chinese Academy of Sciences.

\appendix

\section{Simplified TNMC for the 1D Ising model}
\label{sec:1d}
For the reader's convenience, in this appendix, we first explain some basic concepts and operations 
in the TNMC simulation using the example of the 1D Ising model with free boundary conditions,
for which simple analytical calculations are available and the tensors reduce 
to $2 \times 2$ matrices.
In particular, we shall illustrate the tensor network contraction process and 
the Bayesian sampling process.
Then, we consider the case when the tensor network contractions are approximate and TN samples are biased, and how to use the Metropolis scheme to correct the bias of TN samples.


\subsubsection{Sequential summation and sampling} 
The energy of the 1D Ising model on a chain of $L$ lattice sites with free boundary 
conditions are 
\begin{equation}
E(\mathbf s) = -\sum_{i=2}^L s_{i-1}s_i \; ,
\end{equation}
where the configuration is $\s=\{s_{1},\cdots, s_L\}$, 
and the partition sum $Z$ reads as 
\begin{equation}\label{eq:partition} 
Z = \sum_{s_1=\pm 1}\sum_{s_2=\pm1}e^{ \beta s_1s_2}\cdots\sum_{s_L=\pm1} e^{\beta s_{L-1}s_L} \; .
\end{equation}
In Eq.~(\ref{eq:partition}), the summation over the Ising spins can be 
sequentially taken from the last lattice site $i=L$ to the first site $i=1$.
This {sequential} summation process is analogous to the tensor-network contraction 
process in Fig.~\ref{fig:2d_tn}, and gives rise to a series of process 
functions, $z_i(s_i)$, which play a similar role as those process tensors in Fig.~\ref{fig:2d_tn}.
For instance, by summing over the last Ising spin $s_L$ in Eq.~(\ref{eq:partition}), 
one obtains a bi-value vector $z_i (s_i) =
e^{\beta s_i}+e^{-\beta s_{i}} \equiv 2 \cosh (\beta s_{i})$, 
with $i=L-1$.
With this bi-value vector stored in computer memory, 
the sequential summation process can be expressed as ($i=2,\cdots, L-1$)
\begin{equation} 
z_i (s_{i}) = \sum_{s_{i+1}=\pm 1} 
e^{\beta s_{i} s_{i+1}} z_{i+1} (s_{i+1}) \;  ,
\end{equation}
where $z_L (s_L) =1$ has been set. 
Finally, the total partition function is obtained as $Z= \sum_{s_1=\pm 1} z_1 (s_1)$ 
by summing over the first spin $s_1$.

The sampling of a new spin configuration follows the reverse order of 
the sequential summation process, i.e. from $i=1$ to $L$. 
The first spin is sampled with probability spin $P(s_1) = z_1(s_1)/Z$,
and the first spin $s_1$ being fixed at a specific value $h_1$.   
Then the second spin $s_2$ is sampled with probability $P(s_2|\s_{<2}) = e^{\beta h_1 s_2} \; 
z_2 (s_2)/z_1(h_1)$, where the fixed spin $h_1$ acts like an external field to $s_2$.
Similarly, the probability for the remaining spin can be written as ($i=1,\cdots, L-1$)
\begin{equation}
\label{onedprob} 
P(s_{i+1}|\s_{<{i+1}}) = \frac{ e^{\beta h_i s_{i+1}} \; z_{i+1} (s_{i+1})}{z_i(s_i=h_i)} \; .
\end{equation}
Note that, in such a simple case, 
all the process vectors are constants as 
$z_i (s_i) = (2 \cosh \beta)^{L-i} $ for $ i = 1,\cdots,L$, 
and the total partition function is simply $Z=2 (2 \cosh \beta)^{L-1}$.

\begin{figure}[t]
    \includegraphics[width = 0.70\linewidth]{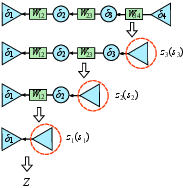}
    \caption{Illustration of the sequential summation process using the language of tensor networks, demonstrated with a 4-spin chain example. The two blue triangles represent the left and right vectors. The contraction proceeds sequentially from $\delta_4$ to $\delta_1$, ultimately yielding the total partition function $Z$. During the contraction process, the tensors enclosed within the red dashed lines correspond to the process vector $z_i(s_i)$.}
    \label{fig:oned}
    \end{figure}

We can see that the overall probability of a new spin configuration 
obeys the Boltzmann distribution as:
\begin{equation}
P(\mathbf s) = \prod_{i=1}^L P(s_i|\s_{<i}) = \frac{1}{Z} e^{-\beta E (\mathbf s)}.
\label{Eq:Poned}
\end{equation}
As illustrated in Fig.~\ref{fig:oned}, the sequential summation processes can be reformulated using the language of tensor networks. In this approach, the copy tensors $\delta_1$ and $\delta_4$ on the boundary lattice sites are reduced to the left vector $(1,1)$ and right vector $\left( \begin{array}{c} 1 \\ 1 \end{array} \right)$. The process vector $z_i(s_i)$ is sequentially obtained by contracting the copy tensor $\delta_{i+1}$ and the Boltzmann matrices $W_{i,i+1}$. The total partition function $Z$ is then obtained by contracting the copy tensor $\delta_1$ with the process vector $z_1(s_1)$. Subsequently, the spins can be sampled sequentially from $s_1$ to $s_4$ based on the probabilities $P(s_i|\s_{<i})$. During the sampling process, after each spin $s_i$ is sampled, the copy tensor $\delta_i$ at that site is reduced to either the left vector $(1,0)$ or $(0,1)$, depending on the spin value, which acts as an 
external magnetic field for the to-be-sampled spin $s_{i+1}$. This reduced vector is then contracted with the Boltzmann matrix $W_{i,i+1}$ and the process vector $z_{i+1}(s_{i+1})$ to calculate the sampling probability $P(s_{i+1}|\s_{<i+1})$ according to Eq.~(\ref{onedprob}). \\

\subsubsection{Metropolis scheme} 
Next, as an illustrative example, we consider the case of periodic boundary conditions. We continue to use the sequential summation and Bayesian sampling approach for free boundary conditions to sample the model with periodic boundary conditions, which is not exact in this case (to make the contraction exact we need to increase the bond dimension). As a consequence, a spin configuration $\s$ under periodic boundary conditions obtained is this way is biased. The energy of the configuration $\s$ with periodic boundary conditions denoted as $E_{\text{P}}(\s)$, and the relationship between $E_{\text{P}}(\s)$ and the energy under free boundary conditions $E(\s)$ is given by $E_{\text{P}}(\s)-E(\s) = - s_1s_L$.

To correct this bias, the Metropolis scheme is introduced, satisfying the detailed balance condition:
\begin{equation} 
    e^{-\beta E_{\text{P}}(\s)}g(\s'|\s)p_a(\s'|\s) = e^{-\beta E_{\text{P}}(\s')}g(\s|\s')p_a(\s|\s'),
\end{equation}
where $g(\s'|\s)$ is the proposal probability to generate a candidate configuration $\s'$ from the current configuration $\s$, and $p_a(\s'|\s)$ is the acceptance probability from configuration $\s$ to $\s'$. Since all candidate configurations are sampled independently, the proposal probability $g(\s'|\s)$ is the $P(\s)$ given in Eq.~(\ref{Eq:Poned}). The candidate configuration $\s'$ is then accepted with probability 
\begin{align}
p_a(\s'|\s)&=\min\left\{1,\frac{P(\s)}{P(\s')}\times\frac{e^{\beta E_{\text{P}}(\s)}}{e^{\beta E_{\text{P}}(\s')}}\right\}\nonumber \\
&=\min\left\{1,\frac{e^{\beta s_1's_L'}}{e^{\beta s_1s_L}}\right\}.
\label{eq:accptanceoned}
\end{align}
This process ensures that, despite the bias introduced by the approximate sampling, the Metropolis scheme allows the system to converge towards the target probability distribution.

\bibliography{references.bib}
\end{document}